\pdfoutput=1
\documentclass[a4paper]{jpconf}
\usepackage{graphicx}
\usepackage{amsmath}
\usepackage{amssymb}
\usepackage{latexsym}
\usepackage{wrapfig}
\begin{document}

\bibliographystyle{iopart-num}

\graphicspath{{images/}}
\DeclareGraphicsExtensions{.pdf}
\pdfimageresolution 110

%#!pdflatex main.tex
% Tai Sakuma <sakuma@fnal.gov>

%%____________________________________________________________________________||
\newcommand{\vecMET}{{\hspace{0.0ex}\not\hspace{-0.8ex}\vec{E}}_\textrm{T}}

\newcommand{\qT}{\vec{q}_\textrm{T}}
\newcommand{\uT}{\vec{u}_\textrm{T}}
\newcommand{\upar}{u_\parallel}
\newcommand{\uper}{u_\perp}

\newcommand{\pT}{p_\text{T}}
\newcommand{\ET}{E_\text{T}}
\newcommand{\Mjj}{M_{\text{jj}}}

\newcommand{\C}{{\cal C}}
\renewcommand{\L}{{\cal L}}
\newcommand{\dt}{\text{d}t}

\newcommand{\pthat}{\hat{p}_\text{T}}
\newcommand{\shat}{\hat{s}}
\newcommand{\costh}{\cos\hat{\theta}}
\newcommand{\GeV}{\text{GeV}}

%%____________________________________________________________________________||

\setlength{\unitlength}{1mm}

\clubpenalty = 10000
\widowpenalty = 10000

%%____________________________________________________________________________||
\title{Missing $E_\text{T}$ Reconstruction with the CMS Detector}

%%____________________________________________________________________________||
\author{Tai Sakuma for the CMS collaboration}

\address{Mitchell Institute for Fundamental Physics and Astronomy \\
Department of Physics \& Astronomy, Texas A\&M University, College
Station, TX 77843, USA }

\ead{sakuma@fnal.gov}

%%____________________________________________________________________________||
\begin{abstract}
The CMS experiment uses missing $E_\text{T}$ to both measure processes
in the Standard Model and test models of physics beyond the Standard
Model. These proceedings show the performance of the missing
$E_\text{T}$ reconstruction evaluated by using 4.6 fb${}^{-1}$ of
proton-proton collision data at the center-of-mass energy $\sqrt{s}$ = 7
TeV collected in 2011 with the CMS detector at the Large Hadron
Collider. Missing $E_\text{T}$ was reconstructed based on a
particle-flow technique. Jet energy corrections were propagated to
missing $E_\text{T}$. After anomalous signals and events were addressed,
the missing $E_\text{T}$ spectrum was well reproduced by MC simulation.
The multiple proton-proton interactions in a single bunch crossing,
pile-up events, degraded the performance of the missing $E_\text{T}$
reconstruction. Mitigations of this degradation have been developed.
\end{abstract}

%%____________________________________________________________________________||
\section{Introduction}

In the CMS experiment, \textit{missing} $E_\text{T}$ (MET) in
proton-proton collisions is reconstructed and used in a wide range of
physics analyses. MET is the imbalance in the transverse momentum of all
\textit{visible particles}, particles which interact with the
\textit{electromagnetic} or \textit{strong forces}, in the final state
of proton-proton collisions. Because momentum is conserved in each
direction, MET is the transverse momentum that must have been carried by
something \textit{invisible}. \textit{Neutrinos}, for example, are
invisible particles; therefore, MET is an estimate of transverse
momentum of neutrinos. We use MET in measurements of \textit{W bosons},
\textit{top quarks}, and \textit{tau leptons} as these particles can
decay into neutrinos. Further, many models of \textit{physics beyond the
Standard Model} predict the existence of particles or something else
which are invisible and can carry momentum; e.g., \textit{Dark Matter}
models, \textit{supersymmetric} models, \textit{unparticle} models, and
models with \textit{large extra dimensions}. For this reason, we use MET
to test such models.

Accurate reconstruction of MET is demanding because it entails
reconstruction of all visible particles in an event with precision. This
requires a \textit{hermetic} detector which can detect all particles
which electromagnetically or strongly interact with matter. The CMS
detector, located at one of two \textit{high luminosity interaction
points} of the \textit{Large Hadron Collider} (LHC), meets this
requirement. The subsystems of the CMS detector include \textit{highly
granular electromagnetic calorimeters}, \textit{hermetic hadronic
calorimeters}, \textit{redundant muon systems}, and \textit{all silicon
trackers} in a \textit{strong magnetic field}. A thorough description of
the CMS detector can be found in Ref.\ \cite{Chatrchyan:2008aa}.

Based on 36 pb${}^{-1}$ of data collected at the center-of-mass energy
$\sqrt{s}$ = 7 TeV in 2010, the CMS collaboration published the results
of comprehensive studies of the MET reconstruction performance
\cite{Chatrchyan:2011tn}. In 2011, the CMS detector collected a
considerably larger amount of data. These proceedings summarize
preliminary results of the MET reconstruction performance study based on
the data collected in 2011 \cite{CMS-DP-2011-010, CMS-DP-2012-003}.

With the increase in the LHC luminosity, the number of multiple
proton-proton interactions in the same \textit{bunch crossing}
(\textit{in-time pile-up}) and overlapping detector signal from previous
or following bunch crossings (\textit{out-of-time pile-up})
significantly increased. The pile-up events worsen the performance of
the MET reconstruction. We have developed several techniques to mitigate
the effect of pile-up events.

\section{The 7 TeV proton-proton collision run in 2011}

\begin{figure*}[!h]
\centering

\includegraphics[scale=0.5]{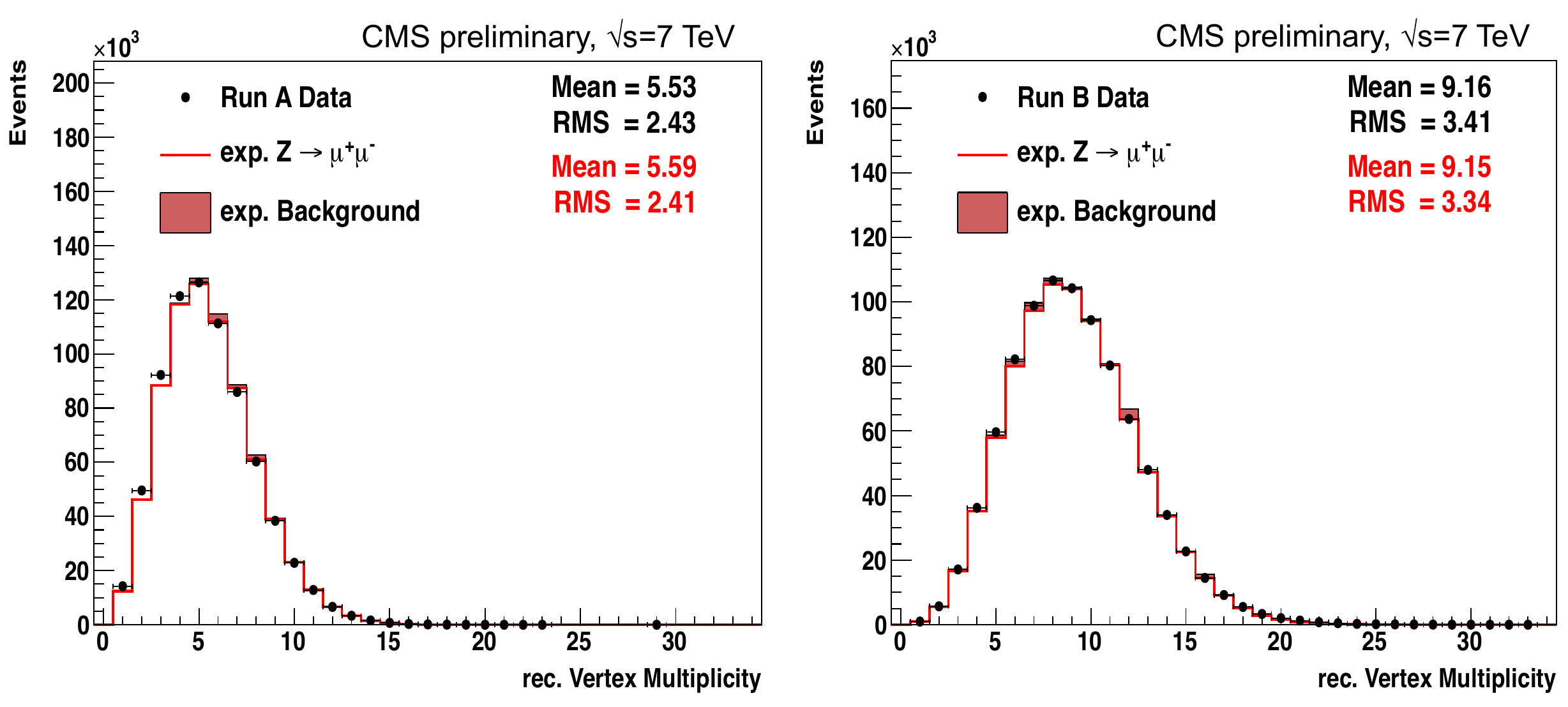}

\caption{\small The distributions of the number of reconstructed
vertices in a single bunch crossing in Run 2011A (left) and Run 2011B
(right). The events contained Z bosons decaying into dimuons. The detail
of the reconstruction and selection of events can be found in Ref.
\cite{CMS-DP-2012-003}.}

\label{143737_22Jul12}
\end{figure*}

These proceedings use 4.6 fb${}^{-1}$ of proton-proton collision data
collected with the CMS detector at the center-of-mass energy $\sqrt{s}$
= 7 TeV in 2011. We divided the data collection period into two
\textit{eras}: Run 2011A and Run 2011B. For proton-proton collisions,
Run 2011A started in March 2011 and ended in August 2011; Run 2011B
started in September 2011 and ended in October 2011. Of the 4.6
fb${}^{-1}$ of data, 2.0 fb${}^{-1}$ was collected in Run 2011A and 2.6
fb${}^{-1}$ was collected in Run 2011B. Since the LHC luminosity rapidly
increased over the year, Run 2011B was shorter than Run 2011A but more
data were collected. In addition, the number of pile-up events also
increased. The average number of the reconstructed \textit{vertices} in
a single bunch crossing, which indicates the average number of in-time
pile-up events, increased from 5.5 in Run 2011A to 9.2 in Run 2011B. The
distributions of the number of the reconstructed vertices in a single
bunch crossing are shown in Figure \ref{143737_22Jul12}.

%%____________________________________________________________________________||
\section{MET reconstruction algorithms and corrections}

MET used in these proceedings is based on a \textit{particle-flow
algorithm} and includes the \textit{Type-I} MET \textit{correction}, a
propagation of \textit{jet energy corrections}. This MET, both with and
without the correction, is called \textit{particle-flow} MET (pfMET) in
Ref.\ \cite{Chatrchyan:2011tn}. Ref.\ \cite{Chatrchyan:2011tn} contains
more detailed description of pfMET and the Type-I MET correction as well
as other definitions of MET and corrections to MET.

First, \textit{raw} MET was defined as the imbalance in the
\textit{transverse energy}\footnote{We will use the transverse momentum
instead of the transverse energy in the near future.} of all particles
in the event reconstructed by a particle-flow algorithm. The detail of
the particle-flow algorithm at CMS can be found in Ref.\
\cite{CMS-PAS-PFT-09-001}. In short, the particle-flow algorithm uses
information from all CMS detector subsystems, i.e., trackers,
calorimeters, and muon systems; then, it reconstructs four momenta of
all visible particles, each of which is identified as one of five
particle types, i.e., \textit{electrons}, \textit{photons},
\textit{muons}, \textit{charged hadrons}, and \textit{neutral hadrons}.

The raw MET is systematically different from \textit{true} MET, i.e.,
the transverse momentum carried by invisible particles, for many reasons
including the non-compensating nature of the calorimeters. To make MET a
better estimate of true MET, the Type-I MET correction was applied. The
Type-I MET correction replaces the transverse energy of particles which
can be clustered as jets with the transverse momentum of the jets to
which jet energy corrections are applied. The detail of jet energy
corrections at CMS can be found in Ref.\ \cite{Chatrchyan:2011ds}.

\section{False MET and event cleaning}

Large MET is caused not only by interesting physics processes in
proton-proton collisions such as production of invisible particles. In
fact, large MET has more often uninteresting causes such as detector
noise, \textit{cosmic rays}, and \textit{beam-halo} particles. MET with
uninteresting causes is called \textit{false} MET or \textit{anomalous}
MET. For an accurate reconstruction of MET, it is, therefore, not
sufficient to reconstruct all visible particles produced in
proton-proton collisions.

\begin{figure*}[!h]
\centering
\includegraphics[scale=0.39]{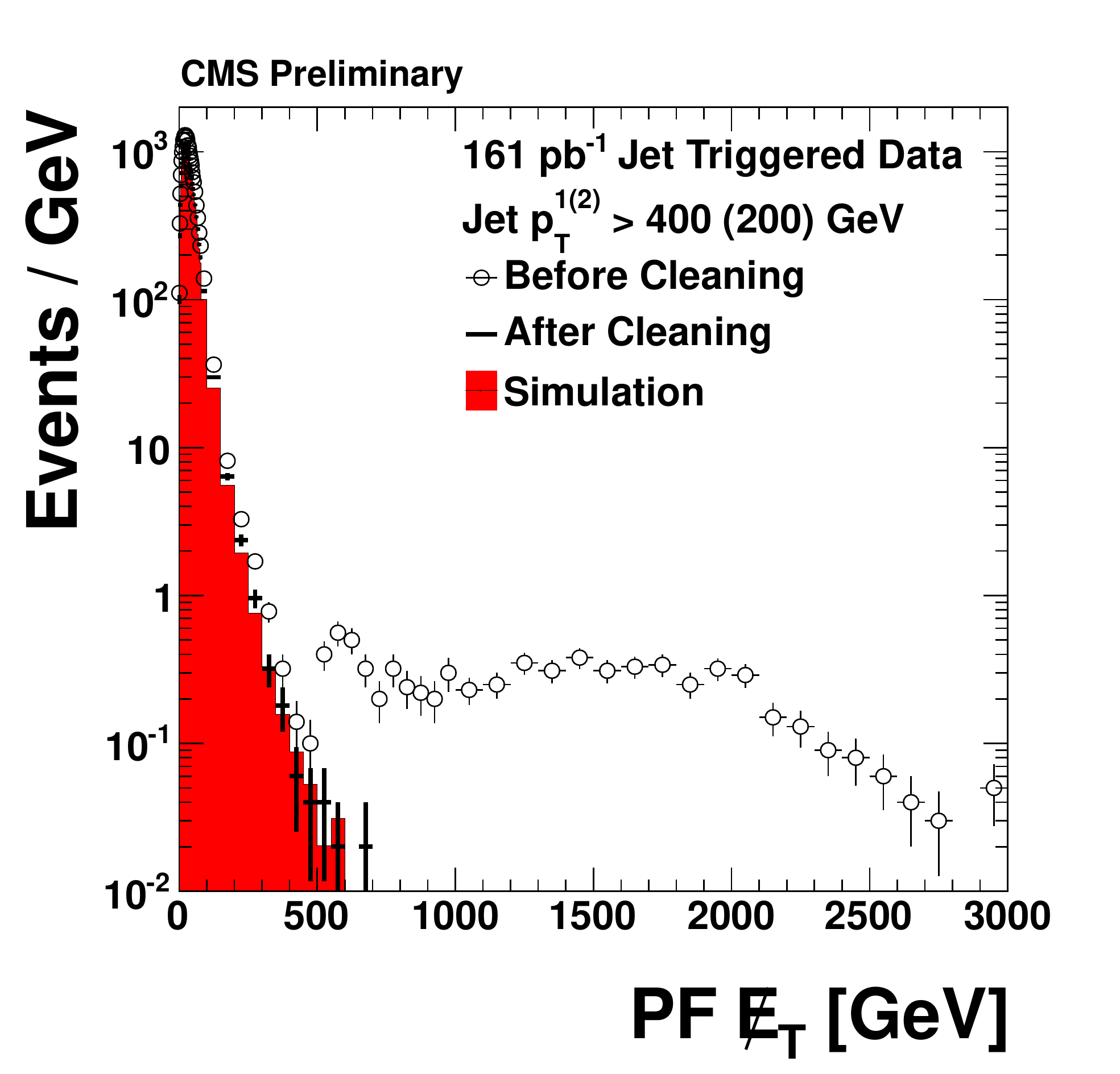}

\caption{\small MET spectrum in dijet events collected in early 2011
before and after the cleaning. The highest bin includes the overflow.
The simulation includes QCD events, top pair production events, W + jets
production events, and Z + jets production events. The simulation does
not explicitly include sources of false MET such as cosmic rays or
beam-halo particles. The detail of the event selection and cleaning can
be found in Ref. \cite{CMS-DP-2011-010}.}

\label{135414_22Jul12}
\end{figure*}

We developed several algorithms to identify false MET. These algorithms,
for example, use timing, pulse shape, and topology of signal. After the
identified false MET was removed, the agreement of the MET spectrum with
simulation, in which causes of false MET were not explicitly simulated,
significantly improved (Figure \ref{135414_22Jul12}).

\section{Performance of MET reconstruction }

MET \textit{response} and \textit{resolution} are measures of MET
reconstruction performance. We evaluated them by artificially inducing
MET in events very likely to have only little or no true MET. We used
events with \textit{vector bosons}, either \textit{Z bosons} decaying
into \textit{dimuons} or photons. These vector bosons are predominantly
produced in interactions with no true MET, such as $qg \rightarrow
q\gamma$, $q\bar{q} \rightarrow Z$, $qg \rightarrow qZ$, $q\bar{q}
\rightarrow gZ$. Therefore, an event is primarily composed of a vector
boson and its \textit{hadronic recoil}. The dimuons and photons were
measured with precision by, respectively, the trackers and muon systems,
and the electromagnetic calorimeters. To induce MET for the performance
evaluation, we excluded the dimuons or photons from the MET
reconstruction; the dimuons or photons played the role of invisible
particles. Then, the MET performance can be indicated by how close
reconstructed MET is to the transverse momentum of the excluded dimuons
or photons.

\begin{wrapfigure}[18]{r}{7.0cm}
\includegraphics[scale=0.35]{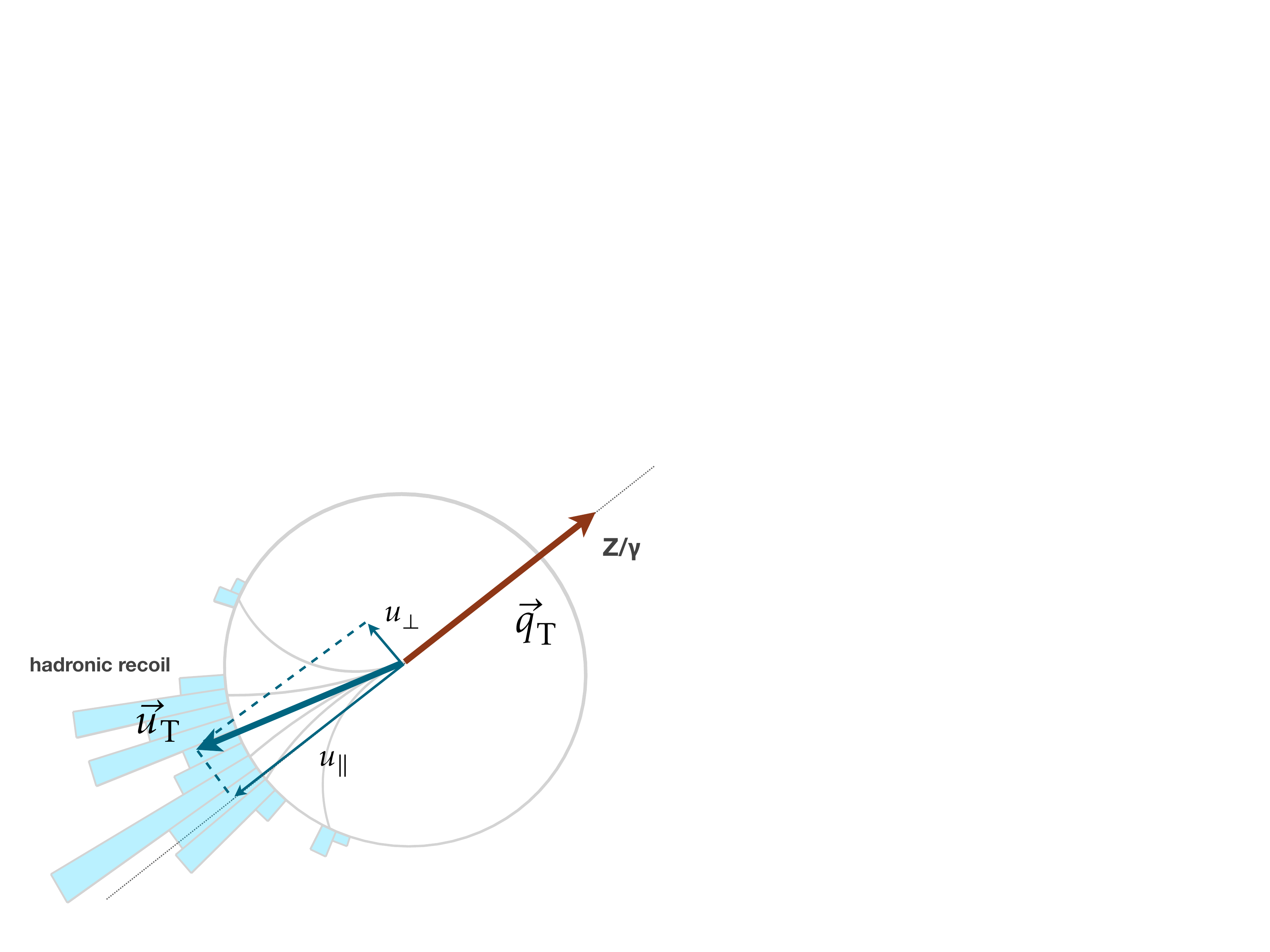} \caption{\small The
transverse momentum of the vector boson and the transverse energy of its
hadronic recoil on the plane perpendicular to the beam axis.}
\label{104058_27Jul12}
\end{wrapfigure}

The transverse momentum of the vector boson is $\qT$. The vector sum of
the transverse energy of all reconstructed particles except the vector
boson, i.e. the hadronic recoil, is $\uT$. Thus, the induced MET is
$\vecMET=-\uT$. The projection of $\uT$ on $\qT$ is $\upar$, which is
typically negative. The component of $\uT$ perpendicular to $\qT$ is
$\uper$. These kinematic variables are illustrated in Figure
\ref{104058_27Jul12}.

\begin{figure*}[!b]
\centering
\includegraphics[scale=0.7]{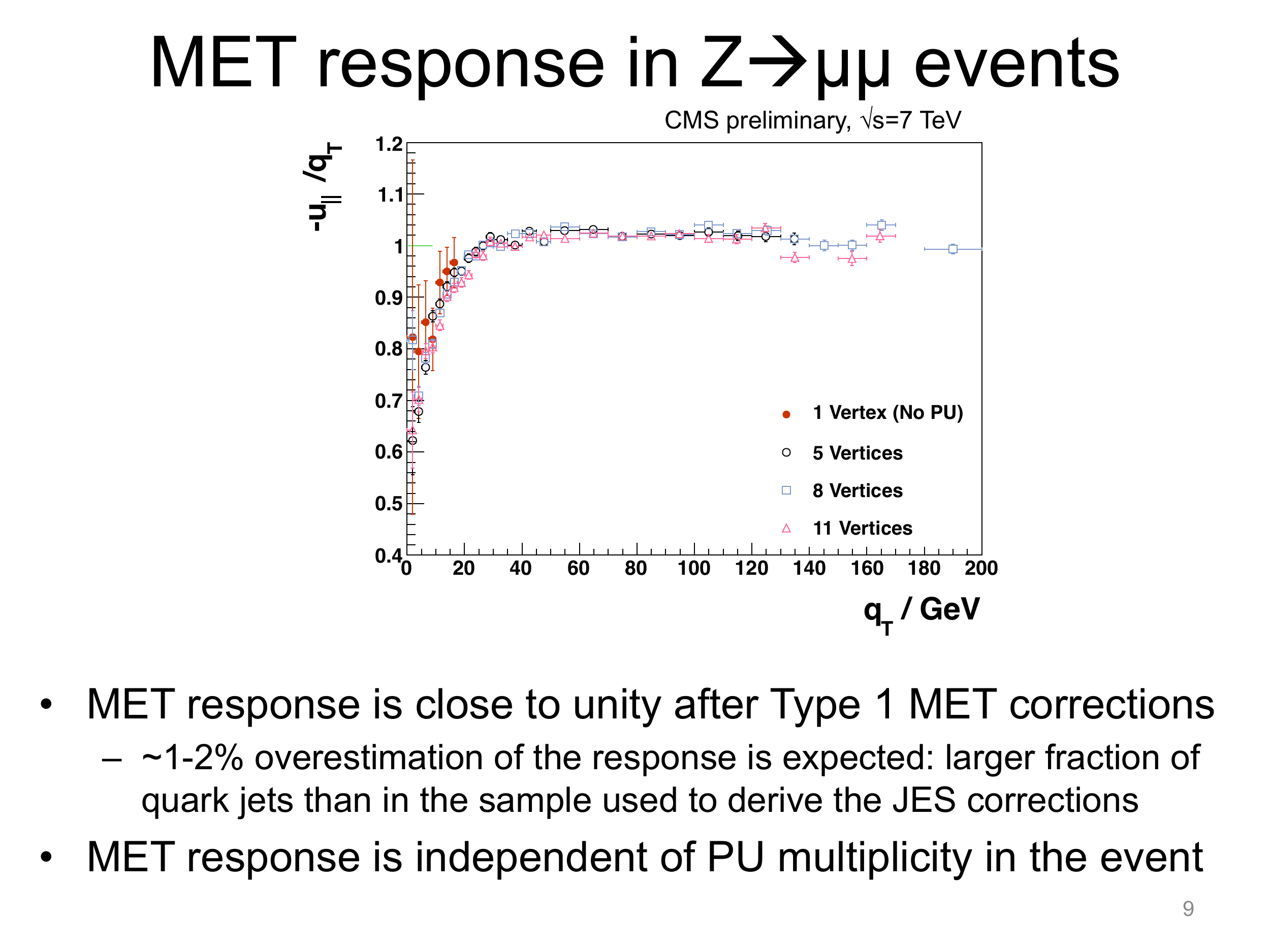}

\caption{\small MET response ($-\upar/|\qT|$) as a function of $|\qT|$
in events with Z bosons decaying into dimuons. Z bosons with transverse
momenta $\qT$ were excluded in MET reconstruction. The projection of
$\vecMET$ on to $\qT$ equals $-\upar$. Events with four different
numbers of vertices are shown as different markers. The detail of the
reconstruction and selection of events can be found in Ref.
\cite{CMS-DP-2012-003}}

\label{105714_27Jul12}
\end{figure*}

MET response is defined as $-\upar/|\qT|$, the ratio of the MET
component parallel to the transverse momentum of the vector boson and
the magnitude of the transverse momentum of the vector boson. Figure
\ref{105714_27Jul12} shows the MET response in events with Z bosons
decaying into dimuons. The response is close to unity at above certain
$|\qT|$ and has little dependence on the number of pile-up events.

\begin{figure*}[!h]
\centering
\includegraphics[scale=0.65]{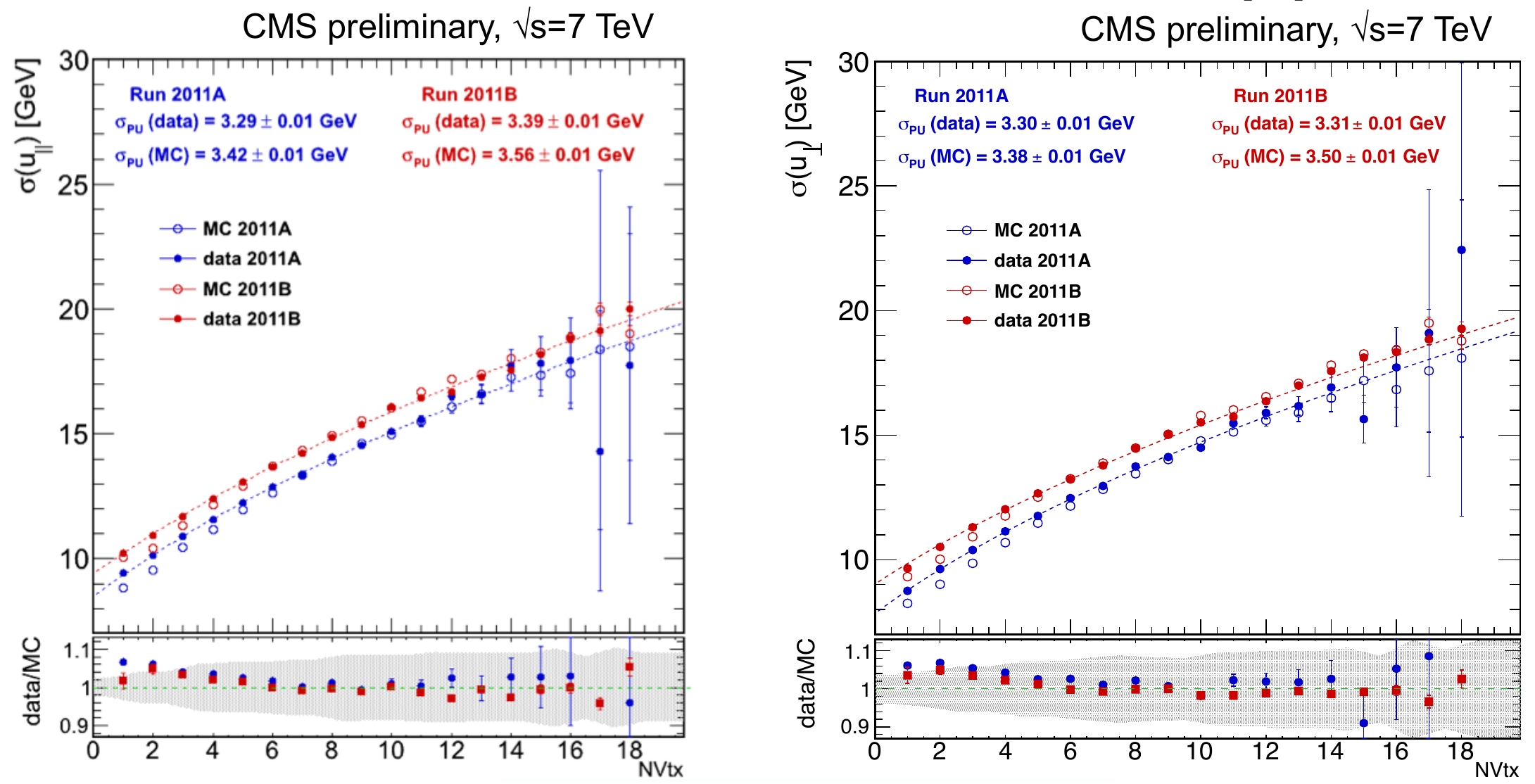}

\caption{\small MET resolutions $\textrm{RMS}(\upar)$ (left) and
$\textrm{RMS}(\uper)$ (right) as functions of the number of
reconstructed vertices ($N_\text{vtx}$) in events with Z bosons decaying
into dimuons. MET resolutions are separately shown for data collected in
Run2011A and Run2011B and for MC simulation for each era. The detail of
the reconstruction and selection of events can be found in Ref.
\cite{CMS-DP-2012-003}}

\label{110832_27Jul12}
\end{figure*}

\begin{figure*}[!h]
\centering

\includegraphics[scale=0.35]{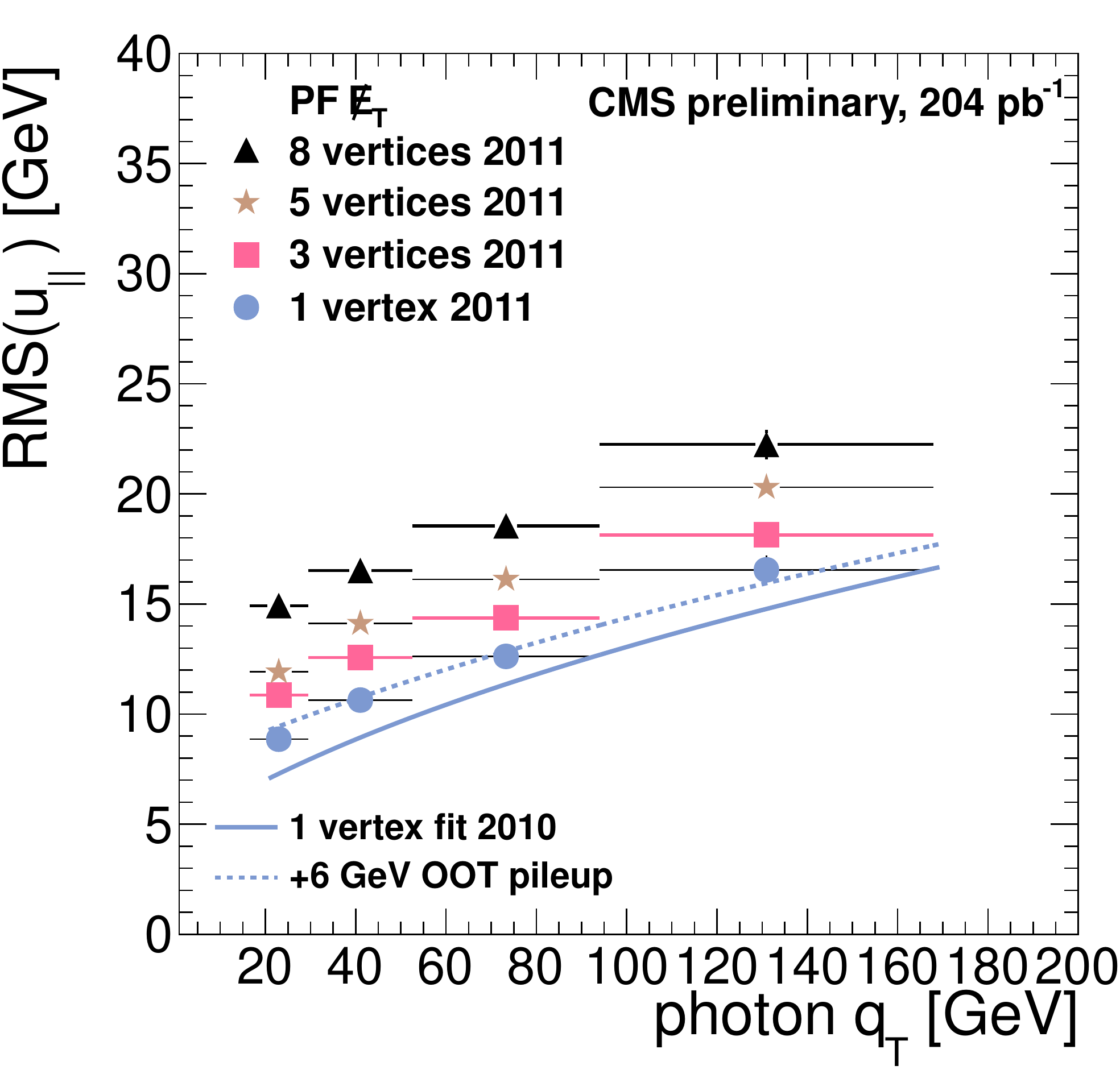}
\includegraphics[scale=0.35]{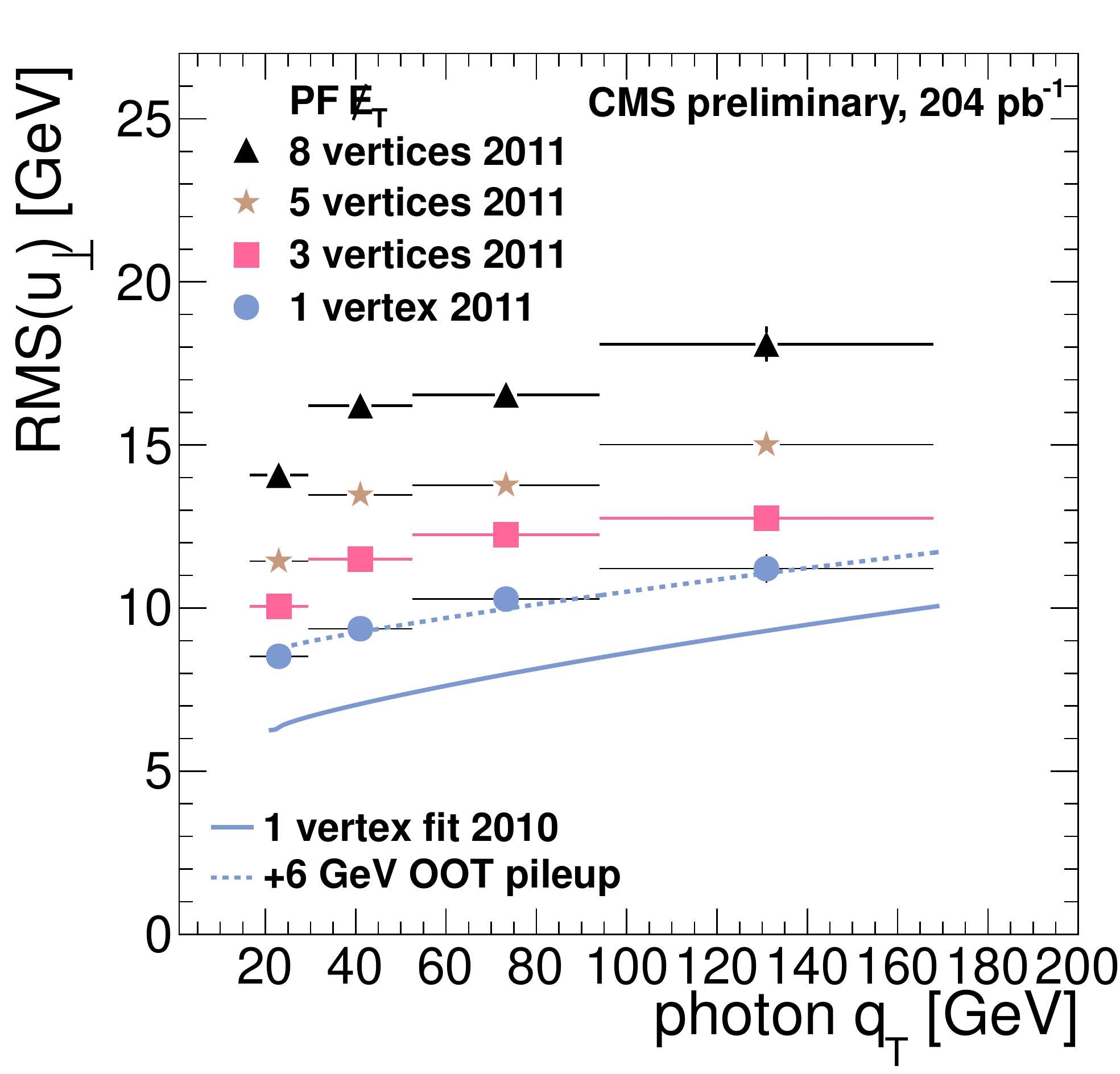}

\caption{\small MET resolutions $\textrm{RMS}(\upar)$ (left) and
$\textrm{RMS}(\uper)$ (right) as functions of the photon $|\qT|$ in
events with photons for data collected in early Run2011A, shown
separately for events with four different numbers of reconstructed
vertices. A curve which fit data for events with one vertex collected in
2010 is also shown. The detail of the reconstruction and selection of
events can be found in Ref. \cite{CMS-DP-2011-010}}

\label{110911_27Jul12}
\end{figure*}

MET resolutions are defined as the RMS of $\upar$ and $\uper$ about
their mean values, denoted, respectively, as $\textrm{RMS}(\upar)$ and
$\textrm{RMS}(\uper)$. Figure \ref{110832_27Jul12} shows the MET
resolutions in Z events as a function of the number of reconstructed
vertices for data collected in Run2011A and Run2011B. Figure
\ref{110911_27Jul12} shows the MET resolutions as functions of $|\qT|$
in photon events for four different numbers of reconstructed vertices
for data collected in early Run2011A. The resolutions degrade as the
number of reconstructed vertices (in-time pile-up events) increases. The
resolutions are worse in events for Run2011B than events with the same
number of in-time pile-up events for Run2011A because the out-of-time
pile-up increased.

\vspace*{5mm}

\vspace*{5mm}

\section{Mitigation of MET degradation in high pile-up events}

This section briefly introduces one of the techniques that CMS used to
mitigate the degradation of the MET performance caused by the pile-up
events.

As shown in the previous section, the MET performance becomes worse as
the number of pile-up events increases. Then, when MET is used in event
selections in physics analyses, the efficiency of selecting signal
events decreases. In order to prevent the efficiency from decreasing, an
analysis of the search for the \textit{Higgs boson} decaying into a pair
of W bosons \cite{Chatrchyan:2012ty} defined another kinematic variable
similar to MET, which is called \textit{TrackMET} in Ref.\
\cite{CMS-DP-2012-003}. TrackMET is the imbalance in the transverse
momentum of charged particles originating from the primary vertex of the
high-$\pT$ event. By its construction, it depends little on the number
of pile-up events. It was shown that using a combination of TrackMET and
MET improved the signal efficiency in this analysis
\cite{Chatrchyan:2012ty}.

%%____________________________________________________________________________||
\section{Summary}

MET is an estimate of the transverse momentum of neutrinos and other
invisible particles or some stuff with momentum. In the CMS experiment,
MET plays a central role in both precision measurements of Standard
Model physics and searches for physics beyond the Standard Model.

Using 4.6 fb${}^{-1}$ of proton-proton collision data at the
center-of-mass energy $\sqrt{s}$ = 7 TeV collected with the CMS detector
in 2011, we evaluated the performance of the reconstruction of MET,
which was based on a particle-flow algorithm, and in particular how it
was affected by pile-up events. After false MET was removed, the MET
spectrum was well described by MC simulation. While the MET response
exhibited little dependence on the number of pile-up events, the MET
resolutions became worse as the number of pile-up events increased. We
have developed several techniques to mitigate the effect of the pile-up
events. As the LHC luminosity is expected to keep rapidly increasing, it
will be important to continue to develop and refine techniques to handle
a large number of pile-up events.

In 2012, the CMS detector is collecting proton-proton collision data at
the center-of-mass energy $\sqrt{s}$ = 8 TeV at the increasing
luminosity. MET will remain an important object to be reconstructed for
a variety of physics analyses at the CMS experiment.

%%____________________________________________________________________________||
\section*{References}
\bibliography{proceedings}

\providecommand{\newblock}{}
\begin{thebibliography}{1}
\expandafter\ifx\csname url\endcsname\relax
  \def\url#1{{\tt #1}}\fi
\expandafter\ifx\csname urlprefix\endcsname\relax\def\urlprefix{URL }\fi
\providecommand{\eprint}[2][]{\url{#2}}

\bibitem{Chatrchyan:2008aa}
{CMS Collaboration} 2008 {\em JINST\/} {\bf 3} S08004

\bibitem{Chatrchyan:2011tn}
{CMS Collaboration} 2011 {\em JINST\/} {\bf 6} P09001 (\textit{Preprint}
  \eprint{1106.5048})

\bibitem{CMS-DP-2011-010}
{CMS Collaboration} 2011 {\em CMS Detector Performance Summary\/} {\bf
  CMS-DP-2011-010}

\bibitem{CMS-DP-2012-003}
{CMS Collaboration} 2012 {\em CMS Detector Performance Summary\/} {\bf
  CMS-DP-2012-003}

\bibitem{CMS-PAS-PFT-09-001}
{CMS Collaboration} 2009 {\em CMS Physics Analysis Summary\/} {\bf
  CMS-PAS-PFT-09-001}

\bibitem{Chatrchyan:2011ds}
{CMS Collaboration} 2011 {\em JINST\/} {\bf 6} P11002 (\textit{Preprint}
  \eprint{1107.4277})

\bibitem{Chatrchyan:2012ty}
{CMS Collaboration} 2012 {\em Phys.Lett.\/} {\bf B710} 91--113
  (\textit{Preprint} \eprint{1202.1489})

\end{thebibliography}

%%____________________________________________________________________________||
\end{document}